\documentclass[reprint,amsmath,amssymb,aps, prl]{revtex4-2}

\usepackage{graphicx}
\graphicspath{{./Plots/}}
\usepackage[space]{grffile}
\usepackage{siunitx}
\usepackage{hyperref}
\usepackage{comment}
\usepackage{tabularx}
\usepackage{bm}
\usepackage{booktabs}

\usepackage{layouts}

\begin{document}
	\title{
			Light-matter interaction at the transition between cavity and waveguide QED
			}
	
	\preprint{APS/123-QED}
	
	\author{Daniel Lechner}
	\affiliation{Department of Physics, Humboldt-Universität zu Berlin, 12489 Berlin, Germany}
	\email{daniel.lechner@physik.hu-berlin.de}
	\author{Riccardo Pennetta}%
	\affiliation{Department of Physics, Humboldt-Universität zu Berlin, 12489 Berlin, Germany}
	\author{Martin Blaha}%
	\affiliation{Department of Physics, Humboldt-Universität zu Berlin, 12489 Berlin, Germany}
	\author{Philipp Schneeweiss}%
	\affiliation{Department of Physics, Humboldt-Universität zu Berlin, 12489 Berlin, Germany}
	\author{Arno Rauschenbeutel}%
	\affiliation{Department of Physics, Humboldt-Universität zu Berlin, 12489 Berlin, Germany}%
	\author{Jürgen Volz}%
	\affiliation{Department of Physics, Humboldt-Universität zu Berlin, 12489 Berlin, Germany}

	\begin{abstract}		
		Experiments based on cavity quantum electrodynamics (QED) are widely used to study the interaction of a light field with a discrete frequency spectrum and emitters.
		More recently, the field of waveguide QED has attracted interest due to the strong interaction between propagating photons and emitters that can be obtained in nanophotonic waveguides, where a continuum of frequency modes is allowed.
		Both cavity and waveguide QED share the common goal of harnessing and deepening the understanding of light-matter coupling.
		However, they often rely on very different experimental set-ups and theoretical descriptions. 
		Here, we experimentally investigate the transition from cavity to waveguide QED with an ensemble of cold atoms that is coupled to a fiber-ring resonator, which contains a nanofiber section. By varying the length of the resonator from a few meters to several tens of meters, we tailor the spectral density of modes of the resonator while remaining in the strong coupling regime. We demonstrate that for progressively longer resonators, the paradigmatic Rabi oscillations of cavity QED gradually vanish, while non-Markovian features reminiscent of waveguide QED appear.
	\end{abstract}
	\maketitle	

\begin{figure}[]
	\centering
	\includegraphics[width = \linewidth]{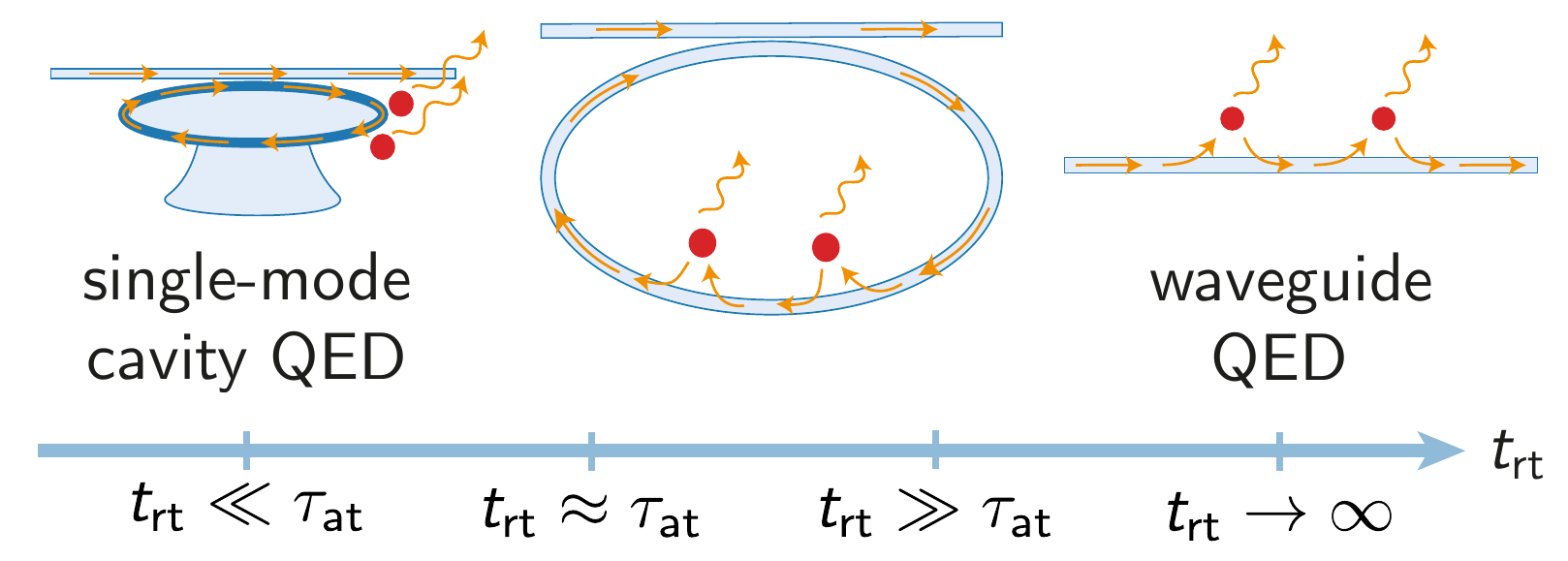}
	\caption{Comparing the resonator roundtrip time, $t_\mathrm{rt}$, with the lifetime of the emitter, $\tau_\mathrm{at}$, allows to distinguish between different regimes of cavity QED. The case of waveguide QED corresponds to a resonator with infinite length and $t_\mathrm{rt} \rightarrow \infty$. To illustrate ring-resonators operating in the different regimes, we show quantum emitters coupled to a whispering-gallery-mode resonator, a fiber-ring resonator and a waveguide.}
	\label{fig:regimes}
\end{figure}
	
	The conventional approach to a quantum description of light-matter interaction has its foundation in the Jaynes-Cummings (JC) model, that considers a single quantum emitter that is coupled to a single-mode optical cavity; the textbook example of cavity quantum electrodynamics (QED) \cite{Jaynes1963, Haroche1989, Gerry2005, Reiserer2015}. In this context, the so-called strong coupling regime is of particular importance. It is reached when the coupling rate between the resonator field and the quantum emitter is much larger than all other decay rates. Under these circumstances, coupling of light and matter can be observed in the frequency domain as a splitting of the resonator mode, known as vacuum Rabi splitting \cite{Thompson1992}. In the time domain, this corresponds to a coherent energy exchange between the quantum emitter and the cavity field, usually referred to as Rabi oscillations \cite{Brune1996, Mielke1997}.	
	Since the early days of quantum optics, the JC model as well as its extension for ensembles of emitters, the Tavis-Cummings (TC) model \cite{Tavis1968, Tavis1969}, have been successfully used to describe a number of different quantum emitters coupled to cavities, including atoms \cite{Rempe1987, Thompson1992, Brune1996}, Bose-Einstein condensates \cite{Brennecke2007, Colombe2007}, molecules \cite{Pscherer2021}, quantum dots \cite{Lodahl2015} and superconducting qubits \cite{Fink2008}. Nonetheless, most optical cavities support not one, but infinitely many frequency modes. Recently, it has been pointed out that the JC and TC models implicitly rely on the assumption that the resonator free spectral range, $\nu_\mathrm{FSR}$, is by far the largest frequency in the system and, in particular, that it is much larger than the natural linewidth of the emitter, $2 \gamma$ \cite{Blaha2022}. In other words, the resonator roundtrip time, $t_\mathrm{rt}=1/\nu_\mathrm{FSR}$, has to be much shorter than the excited-state lifetime of the emitter, $\tau_\mathrm{at}=1/( 2\gamma)$.
	If this assumption is not fulfilled, some distinctive features of cavity QED can be substantially modified. For instance, in the limit when the distance between the cavity mirrors approaches infinity, the emitters interact with a continuum of frequency modes. Now, this scenario is investigated in the rapidly growing field of waveguide QED, which considers single-pass interactions between an optical mode that propagates through a waveguide and ensembles of quantum emitters \cite{Chang2018, Sheremet2022}. In this context, phenomena such as the vacuum Rabi splitting disappear and the light-matter coupling strength is rather measured by the ensemble's optical depth (OD).
	
	Despite its conceptual and practical significance, the transition between cavity and waveguide QED (see Fig.~\ref{fig:regimes}) has only been discussed in a handful of works \cite{Meiser2006, Krimer2014, Blaha2022}. This may be partially due to the fact that most experimental setups employ free-space cavities, in which, due to diffraction, an increase of the cavity length has to be accompanied by an increase of the mirror diameter if one seeks to remain in the strong coupling regime. Practically, this limits the length of most optical cavities to the centimeter range, in which the single mode approach of the JC and TC model usually still holds.
	
	In this Letter, we experimentally explore the transition between cavity and waveguide QED in an all-fiber ring-resonator including a nanofiber-based optical interface for cold atoms. This system allows us to tailor the density of modes in the frequency domain by varying the length of the fiber resonator. We present measurements obtained by coupling resonators with geometric lengths, $L_\mathrm{cav}$, of \SI{5.8}{\meter} and \SI{45.4}{\meter} to an ensemble of cold cesium atoms, probed on the D2 line, for which $\tau_\mathrm{at} = \SI{30.4}{\nano\second}$ \cite{Steck2019}. The two resonator lengths were chosen such that for the former $t_\mathrm{rt} \approx \tau_\mathrm{at}$ and for the latter $t_\mathrm{rt} \gg \tau_\mathrm{at}$. Due to the absence of diffraction and the negligible propagation loss of optical fiber for the length scales considered here, the cavity length in this setup has no influence on the system's cooperativity parameter, $C = \frac{g^2}{2\kappa \gamma}$; where $g$ is the ensemble--resonator coupling rate and $\kappa$ is the cavity loss rate \cite{Reiserer2015, Schneeweiss2017, Ruddell2017}.
	In particular, we concentrate on one of the key features of atom-light interaction in cavity QED, namely the phenomenon of vacuum Rabi oscillations. We experimentally demonstrate that, when increasing the cavity length, Rabi oscillations are progressively replaced by non-Markovian features \cite{Krimer2014, Sinha2020, Ferreira2021} that repeat at multiple integers of the roundtrip time and are reminiscent of the superradiant dynamics of waveguide QED. Eventually, vacuum Rabi oscillations cease to exist in the so-called superstrong coupling or multi-mode strong coupling regime of cavity QED \cite{Meiser2006, Krimer2014,Sundaresan2015, Johnson2019, Blaha2022}, in which several cavity modes simultaneously interact with the atomic ensemble.

		As mentioned before, the JC and TC models correctly capture the dynamics of short resonators (i.e., if $t_\mathrm{rt} \ll \tau_\mathrm{at}$). To theoretically describe light-matter interactions for longer resonators, in which $t_\mathrm{rt}$ becomes comparable or even larger than $\tau_\mathrm{at}$, we follow the approach of Refs. \cite{Shen2009, Blaha2022, Pennetta2022}. There, a more general real-space quantum mechanical approach is developed, which takes the cascaded interaction of light with individual atoms into account and includes coupling to all cavity frequency modes. This allows us to describe the delayed coherent feedback to the atomic ensemble, which occurs in long resonators. In the following, we refer to this theoretical approach as cascaded interaction (CI) model.

\begin{figure}[]
		\centering
		\includegraphics[width = \linewidth]{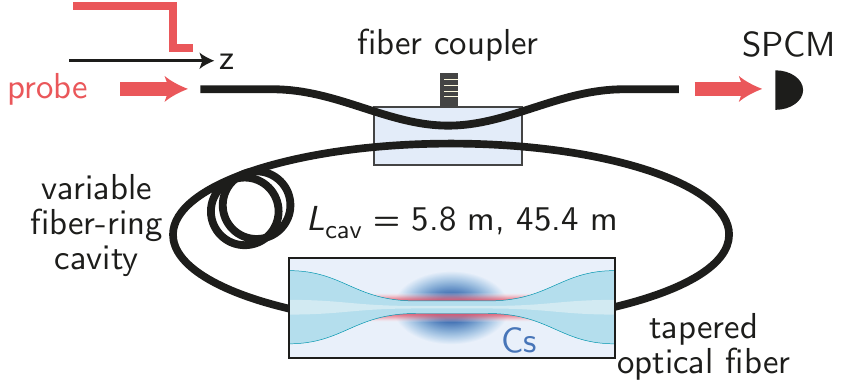}
		\caption{Experimental setup of our fiber-ring cavity system with variable cavity length, $L_\mathrm{cav}$. The ensemble--cavity system is probed in transmission with a single photon counting module (SPCM) after the rapid switch-on of a probe laser resonant with the D2 line of cesium (Cs).}
		\label{fig:Setup}
	\end{figure}

	Our experimental setup is sketched in Fig.~\ref{fig:Setup} and consists of a fiber-ring resonator containing an optical nanofiber section that is realized as the waist of a tapered optical fiber.	
	The nanofiber has a nominal diameter of \SI{400}{\nano\meter} and a nominal length of \SI{10}{\milli\meter}. The evanescent field surrounding the nanofiber allows us to interface the resonator to an ensemble of cold cesium atoms that issues from a magneto-optical trap.
	We use a variable fiber coupler to couple light to and from a ring-resonator. By splicing different lengths of single-mode fiber between the two ends of the fiber coupler, the roundtrip time of the resonator can easily be varied. 
	The time domain response of the system is investigated by recording the transmission through the fiber coupler with a single-photon counting module (SPCM) after the switch-on of a weak probe pulse with a power that integrates to less than one photon per excited state lifetime. By employing an electro-optic modulator in Mach-Zehnder configuration, we measure a rise-time of the pulse of $\approx \SI{850}{\pico\second}$, which is much faster than the lifetime of the excited state of cesium. 
	All experiments are performed with the probe laser as well as the cavity on resonance with the D2 line of cesium $(6^2S_{1/2}, \, F= 4 \rightarrow 6^2P_{3/2}, \, F = 5)$.
	To characterize the steady-state frequency response of the system, we sweep a continuous wave probe laser across the atomic transition and record the transmission with the SPCM.
	\begin{figure*}[]
		\centering
		\includegraphics[width = 1\linewidth]{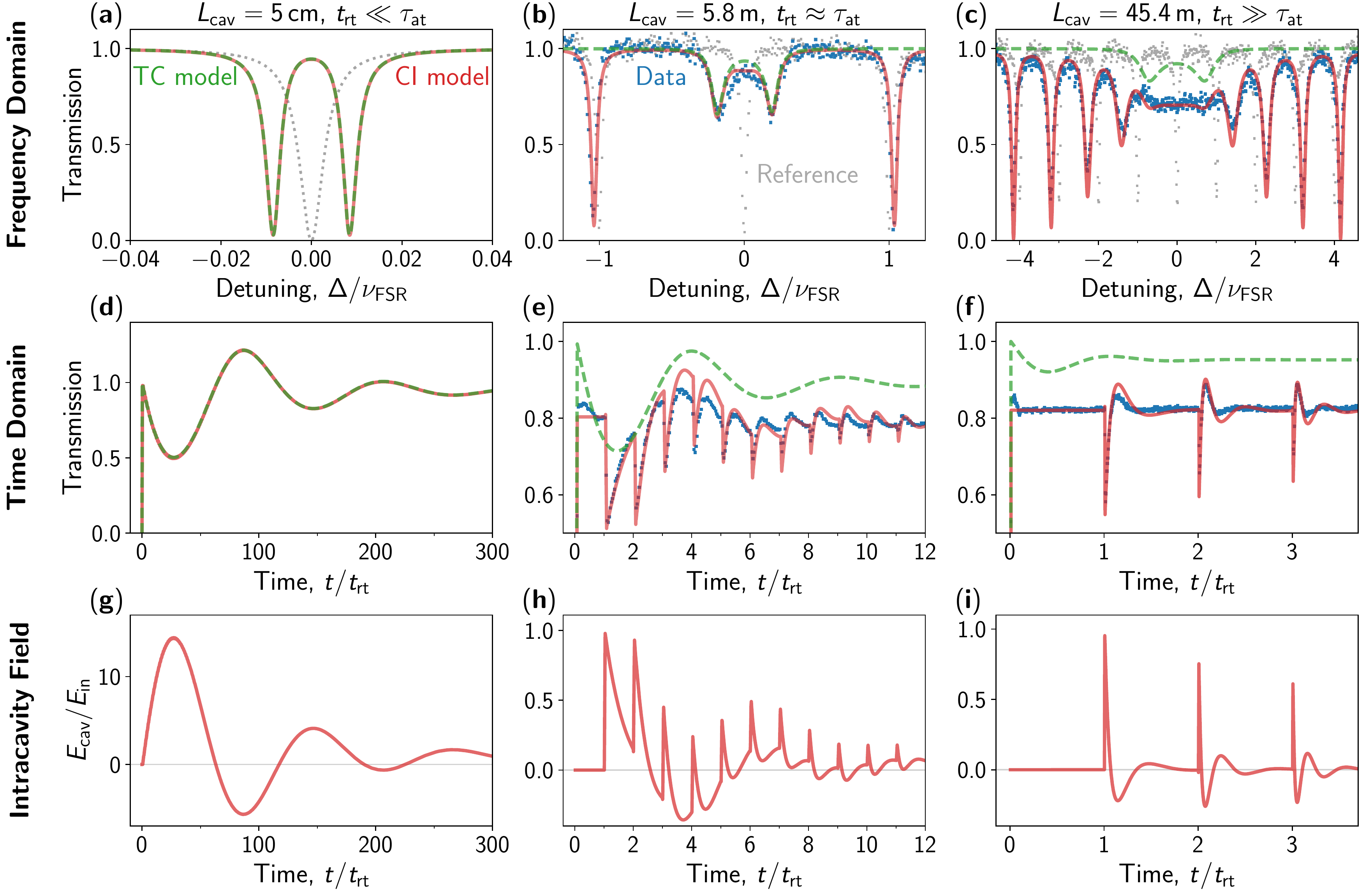}
		\caption{Transmission through the ensemble--cavity system in the frequency domain (top panels, (a, b, c)) and in the time domain (center panels, (d, e, f)) for different resonator lengths, $L_\mathrm{cav}$. Experimental data is represented by blue squares, while the reference spectra for empty cavities are shown in grey. The predictions from the Tavis-Cummings (TC) and the cascaded interaction (CI) model are plotted by green dashed lines and red solid lines, respectively.
		The intermediate regime (b, e) simultaneously shows features from the single-mode (a, d) and the multi-mode case (c, f).
		(g, h, i) Calculated intracavity field amplitudes just before the fiber coupler, $E_\mathrm{cav}(t)$, normalized to the field entering the cavity at $t = 0$, $E_\mathrm{in}$, for the parameters of (d - f), respectively. The ratio $E_\mathrm{cav}(t) / E_\mathrm{in}$ is real-valued because the probe field is resonant with both the cavity and the atomic ensemble. $t_\mathrm{rt}$: resonator roundtrip time, $\tau_\mathrm{at}$: natural lifetime of emitters, $\Delta$: laser-cavity detuning, $\nu_\mathrm{FSR}$: resonator free spectral range.
		} 
		\label{fig:Data}
	\end{figure*}
	The frequency and temporal response of the system for ring-resonators of different lengths is depicted in Fig.~\ref{fig:Data}. For completeness, we show our measurements together with the predictions of the TC and CI models obtained for a much shorter 5-cm-long ring-resonator, operating in the single-mode regime of cavity QED. This regime is not easily accessible with our experimental platform, but has been widely studied in literature \cite{Thompson1992, Brune1996, Bochmann2008}. To adequately compare the different resonators, we chose the system parameters to ensure operations in the strong coupling regime while, at the same time, keeping the ratio between $g$ and the maximum of $\kappa$ and $\gamma$ approximately constant, within the limits of our experimental apparatus (see Tab.~\ref{tab:params}). In the TC model, this ratio indicates the number of Rabi oscillation the system undergoes before reaching a steady state.
{\renewcommand{\arraystretch}{1.15}
	\begin{table*}[t] 
		\caption{Cavity parameters for the three resonators investigated in Fig.~\ref{fig:Data}. For the D2 line of cesium, the natural lifetime of the excited state is $\tau_\mathrm{at} = \SI{30.4(2)}{\nano\second}$ and the spontaneous emission rate is $\gamma = 2\pi \cdot  \SI{2.62(1)}{\mega\hertz}$ \cite{Steck2019}. The errors were estimated from the fit as well as the time resolution of our time domain measurements. The number in parenthesis represents the 95\% confidence interval of the measured quantity.}\label{tab:params}
		\vspace{8pt}
		\begin{tabularx}{16.6cm}{p{6.5cm} p{2.5cm} p{2.5cm}p{2.5cm}p{2.5cm}}
			&& $ \bm{t_\mathrm{rt} \ll \tau_\mathrm{at} }$ & $ \bm{ t_\mathrm{rt} \approx \tau_\mathrm{at} }$ & $ \bm{ t_\mathrm{rt} \gg \tau_\mathrm{at} }$ \\ \toprule 
			cavity length & $L_\mathrm{cav}$ &  5.0 cm & 5.843(5) m &45.423(5) m	\\ 
			
			cavity roundtrip time & $t_\mathrm{rt}$ & 241.67 ps & 28.26(3) ns & 219.70(3) ns \\
			
			free spectral range & $\nu_\mathrm{FSR}$ & 4.14 GHz & 35.38(3) MHz & 4.5517(6) MHz \\
			
			optical depth & OD &1.42 & 6.61(7) & 14.4(1)\\

			ensemble--resonator coupling rate& $g$ & $2\pi \cdot \SI{34.98}{\mega\hertz}$ &  $2\pi \cdot \SI{6.98(4)}{\mega\hertz}$ & $2\pi \cdot \SI{3.70(2)}{\mega\hertz}$ \\
			
			cavity loss rate & $\kappa$ & $2\pi \cdot \SI{13.11}{\mega\hertz}$ & $2\pi \cdot \SI{919.6(3)}{\kilo\hertz}$  & $2\pi \cdot \SI{114.2(1)}{\kilo\hertz}$ \\
			cooperativity & $C$ &17.84 & 10.1(1) & 22.9(2)\\
			ratio between $g$ and dominating loss rate & $g / \mathrm{max}(\kappa,\gamma)$ & 2.67 & 2.67(2) & 1.413(6)\\

		\end{tabularx}
	\end{table*}
}

	The standard case of a single-mode cavity is illustrated in Fig.~\ref{fig:Data}~(a, d, g) in which a 5-cm-long resonator ($t_\mathrm{rt} = \SI{241.7}{\pico\second} \ll \tau_\mathrm{at} = \SI{30.4}{\nano\second}$) is coupled to an atomic ensemble with $\mathrm{OD} = 1.42$. A clear vacuum Rabi splitting can be observed in the frequency domain (Fig.~\ref{fig:Data}~(a)) around zero detuning, $\Delta = 0$. 
	As expected, the time domain (Fig.~\ref{fig:Data}~(d)) is characterized by conventional sinusoidal Rabi oscillations, which are damped due to dissipation.
	To further clarify the time domain signal, we plot in Fig.~\ref{fig:Data}~(g) the calculated ratio between the intracavity field amplitude just before the fiber coupler, $E_\mathrm{cav}(t)$, and the amplitude of the field entering the cavity at $t = 0$, $E_\mathrm{in}$. 
	The observed oscillations of the intracavity field are due to the periodic exchange of energy stored in the cavity and in the atoms. The latter absorb the propagating field in the resonator and, on resonance, re-radiate the field with a phase shift of $\pi$.
	The calculated signal at the detector is shown in Fig.~\ref{fig:Data}~(d) and is the result of the interference of the light transmitted by the fiber coupler, which never enters the cavity, and the fraction of the intracavity field (Fig.~\ref{fig:Data}~(g)) that is outcoupled.
	Due to the cavity resonance condition, these two fields have a $\pi$-phase difference at $t=0$ and the detected signal is an inverted version of the intracavity field.

	The very different scenario of a ring-resonator operating in the superstrong coupling or multi-mode regime of cavity QED is shown in Fig.~\ref{fig:Data}~(c, f, i). Here, we present experimental and theoretical results obtained by coupling an ensemble with $\mathrm{OD} = 14.4 $ to the 45.4-m-long resonator ($t_\mathrm{rt} = \SI{219.7}{\nano\second} \gg \tau_\mathrm{at} = \SI{30.4}{\nano\second}$). The multi-mode nature of light-matter coupling can be clearly observed in the frequency response of the system (see Fig.~\ref{fig:Data}~(c)). Together with the familiar splitting at $\Delta = 0$, several other modes are shifted outwards compared to the empty resonator spectrum \cite{Meiser2006, Johnson2019}. This system clearly operates beyond the validity of the TC model, which only qualitatively reproduces the vacuum Rabi splitting, see green dashed line.
	However, it is in the time domain (see Fig.~\ref{fig:Data}~(f)), that the observed dynamics deviate even more drastically from standard cavity QED. No Rabi oscillations are visible and, instead, the transmission is characterized by sharp features separated by the roundtrip time. This is in very good agreement with the prediction of the CI model, which was obtained by fitting the data with the OD as the only free parameter. All other cavity parameters were independently measured from the empty cavity response.	
	To understand the origin of these sharp features, we consider again the intracavity field illustrated in Fig.~\ref{fig:Data}~(i). The light field enters the cavity at $t = 0$ and reaches the output coupler at $t/t_\mathrm{rt} = 1$, when the switch-on of the laser pulse can be observed. Then, the field amplitude starts to decay due to absorption by the atomic ensemble, which happens on a time scale much faster than the cavity roundtrip time. 
	In particular, taking into account collective light-matter coupling, the decay of the intracavity field amplitude occurs with an initial rate approximately given by $(1 + \mathrm{OD}/4) / (2\tau_\mathrm{at})$ \cite{Pennetta2022a}. This process unfolds unperturbed until the light completes a full cavity roundtrip. Since $t_\mathrm{rt}$ is so long, the light field reaches a steady state with $E_\mathrm{cav} / E_\mathrm{in} = {e}^{-\mathrm{OD} / 2}$, meaning that increasing the cavity length further would no longer qualitatively alter the results. Up to this point, the system response is completely equivalent to the single-pass dynamics of waveguide QED. At $t/t_\mathrm{rt} = 2$, the atoms are again driven by the switch-on of the laser pulse at its second roundtrip. The absorption process repeats now in a similar fashion, however, its initial decay rate is faster.
	This is because the light field that reaches the atoms at the second roundtrip has already interacted with the same ensemble once. Therefore, from a waveguide QED perspective, the observed behavior should be compared to a single-pass through an ensemble with twice the OD \cite{Pennetta2022}. In general terms, the intracavity dynamics after the $m^\mathrm{th}$ roundtrip is approximately the same as a single-pass through an ensemble with a total optical depth $\mathrm{OD}_\mathrm{tot} = m \times \mathrm{OD}$. In fact, in this regime, no build-up of the intracavity field occurs and the ring-resonator acts as a non-Markovian reservoir, meaning that it provides delayed coherent feedback to the atomic ensemble. 
	We note that the oscillations that follow the initial decay of the intracavity field are typical features of collective light-matter coupling for ensembles of emitters arranged in a single spatial dimension \cite{Pennetta2022a}. Also here, when considering the signal at the detector in Fig.~\ref{fig:Data}~(f), all the features that we discussed above appear inverted due to the $\pi$-phase shift that follows from the cavity resonance condition.

	A surprisingly rich dynamics characterizes the intermediate regime, that we reached by coupling the 5.8-m-long ring-resonator ($t_\mathrm{rt} = \SI{28.26}{\nano\second} \approx \tau_\mathrm{at} = \SI{30.4}{\nano\second} $) to an ensemble with $\mathrm{OD} = 6.61$. The frequency response of the system is shown in Fig.~\ref{fig:Data}~(b), where, at the first sight, the multi-mode nature of light-matter coupling seems negligible. Indeed, the observed vacuum Rabi splitting is reasonably well reproduced by the TC model, which, however, incorrectly predicts the steady-state transmission for resonant excitation. Furthermore, the neighboring cavity modes exhibit a modest outwards shift when compared to the empty resonator spectrum.
	However, a careful analysis of the time domain response of the system reveals a clearer picture. Notably, our measurement of the transmitted light (see Fig.~\ref{fig:Data}~(e)) exhibits conventional Rabi oscillations superimposed with the non-Markovian dynamics, i.e., the system simultaneously exhibits single-mode and multi-mode features. To further support this claim, we note that the envelope of the measured signal (i.e., the Rabi oscillations) roughly coincides with the predictions of the TC model.
	The small discrepancies between our measurements and the CI model can be attributed to inhomogeneous broadening in the atomic ensemble.	
	This rather complex behavior can be clarified considering the time evolution of the intracavity field shown in Fig.~\ref{fig:Data}~(h). Here, it is possible to see that the initial dynamics is very similar to the multi-mode case discussed before. Again, the switch-on of the laser pulse at $t/t_\mathrm{rt} = 1$ is followed by a decay due to atomic absorption, at a rate that only depends on $\tau_\mathrm{at}$ and OD. However, since $t_\mathrm{rt}$ is significantly shorter in this case, there is not enough time for the intracavity field to reach the steady state before the first roundtrip is completed. For this reason, roundtrip after roundtrip, the intracavity field starts to build up and Rabi oscillations similar to those depicted for the single-mode case (see Fig.~\ref{fig:Data}~(g)) appear together with the non-Markovian dynamics.
	
	In conclusion, in this Letter we presented a detailed experimental study of collective light-matter coupling at the transition between cavity and waveguide QED. The use of an all-fiber system allows us to experimentally access new regimes of cavity QED, in which the non-Markovian response of the cavity becomes relevant and key features of light-matter coupling, as for instance Rabi oscillations, gradually disappear. 
	We note that the three regimes considered here offer a comprehensive overview of possible scenarios, as increasing or decreasing the resonator length any further does not qualitatively alter the response of the system.
	Observing these effects with cold atomic ensembles required unusually long resonators. However, for other types of quantum emitters, with orders of magnitude larger intrinsic linewidth, these effects become relevant for surprisingly short resonators. For instance, for quantum dots with radiative lifetimes between $\SI{100}{\pico\second}$ to $\SI{1}{\nano\second}$ \cite{Lodahl2015} non-Markovian features should already appear when coupling them to resonators just a few centimeters long.	
	The properties of non-Markovian systems featuring coherent time-delayed feedback have been recently discussed in the context of photonic circuits \cite{Pichler2016} and collective spontaneous emission between distant quantum emitters \cite{Guimond2016a, Sinha2020}. From this perspective, our results shed light on the interplay between the single-pass collective response of the atoms and the resulting ensemble--cavity dynamics and have the potential to unveil novel and unintuitive phenomena in these new regimes of light-matter interaction.
	
	\begin{acknowledgments}
We thank L. A. Orozco for helpful discussions. We acknowledge financial support by the Alexander von Humboldt Foundation in the framework of an Alexander von Humboldt Professorship endowed by the Federal Ministry of Education and Research as well as by the Austrian Science Fund (NanoFiRe grant project No. P31115).

D.~Lechner and R.~Pennetta contributed equally to this work.
\end{acknowledgments}
	
	\bibliography{library.bib}
	
\end{document}